\documentclass[10pt,letterpaper,twocolumn]{article}

\usepackage[letterpaper,
  top=0.98in,
  bottom=0.98in,
  left=0.98in,
  right=0.98in,
  headheight=22pt,
  headsep=8pt,
  footskip=24pt]{geometry}

\usepackage[T1]{fontenc}
\usepackage{lmodern} 
\IfFileExists{newtxtext.sty}{\usepackage{newtxtext,newtxmath}}{}
\usepackage{microtype}
\usepackage{graphicx}
\usepackage{booktabs}
\usepackage{array}
\usepackage{amsmath}
\usepackage{enumitem}
\usepackage{titlesec}
\usepackage{fancyhdr}
\usepackage{lastpage}
\usepackage{caption}
\usepackage{url}
\usepackage[hidelinks]{hyperref}

\usepackage{tikz}
\usetikzlibrary{positioning}

\newcommand{\IACPaperCode}{IAC-26}
\newcommand{\IACTitle}{Towards a GPU-Accelerated Real-Time Imaging Pipeline for SETI\\ at the Allen Telescope Array}
\newcommand{\IACConference}{77\textsuperscript{th} International Astronautical Congress (IAC 2026), Antalya, T\"urkiye, 5--9 Oct 2026.}

\titleformat{\section}
  {\normalfont\bfseries\fontsize{10}{11}\selectfont}
  {\thesection.}{0.45em}{}
\titlespacing*{\section}{0pt}{9pt}{1pt}

\titleformat{\subsection}
  {\normalfont\itshape\fontsize{10}{11}\selectfont}
  {\thesubsection}{0.45em}{}
\titlespacing*{\subsection}{0pt}{7pt}{1pt}

\titleformat{\subsubsection}
  {\normalfont\itshape\fontsize{10}{11}\selectfont}
  {\thesubsubsection}{0.45em}{}
\titlespacing*{\subsubsection}{0pt}{6pt}{1pt}

\newcommand{\iacunnumbered}[1]{%
  \par\vspace{7pt}%
  \noindent\textbf{#1}\par\vspace{1pt}}

\setlist{nosep,leftmargin=*}

\begin{document}

\twocolumn[{%
\begin{minipage}{\textwidth}
\begin{center}
  \fontsize{11}{12}\selectfont

  \vspace{8pt}

  {\bfseries\fontsize{12}{14}\selectfont \IACTitle\ }

  \vspace{8pt}

  {\bfseries\fontsize{11}{13}\selectfont
  Karen I. Perez\textsuperscript{a,b,*}, Joe S. Bright\textsuperscript{a}, Vishal Gajjar\textsuperscript{a,c}, Luigi F. Cruz\textsuperscript{a,d}}
\end{center}

\vspace{5pt}

\noindent\textsuperscript{a}\textit{SETI Institute, 339 Bernardo Ave, Suite 200, Mountain View, CA 94043, USA}\par
\vspace{-1pt}
\noindent\textsuperscript{b}\textit{Steward Observatory, University of Arizona, 933 N Cherry Ave, Tucson, AZ 85719, USA}\par
\vspace{-1pt}
\noindent\textsuperscript{c}\textit{Breakthrough Listen, University of California, Berkeley, 501 Campbell Hall \#3411, Berkeley, CA 94720, USA}\par
\vspace{-1pt}
\noindent\textsuperscript{d}\textit{Berkeley SETI Research Center, UC Berkeley, 339 Campbell Hall, Berkeley, CA, USA}

\vspace{8pt}

\noindent * Corresponding Author: Karen I. Perez; kperez@seti.org

\vspace{11pt}

\begin{center}
\textbf{Abstract}
\end{center}
\vspace{-4pt}

\noindent
Traditional technosignature and transient searches commonly operate on beamformed time--frequency products. Interferometric imaging adds spatial context and allows time-domain searches to be performed across many image pixels simultaneously. We report the development and testing of a GPU-accelerated interferometric imaging and image-domain dispersion-measure (DM) search pipeline for Allen Telescope Array (ATA) post-correlation UVH5 visibility data, integrated with the CyberEther streaming framework to replay recorded visibility data and work toward real-time operation. The pipeline forms Stokes-I dirty images from correlated visibilities, incorporates Inter-Quartile Range Mitigation (IQRM) flags into the imaging weights, images 48 frequency subbands spanning 287.5\,MHz, subtracts a temporal mean over the full observation from each subband and pixel, applies IQRM-informed subband masking, and performs a blind trial-DM search using a short-history image ring buffer. On a controlled 10-minute three-spectral-window test observation containing a periodic dispersed pulsed injection (DM\,$\approx$\,150\,pc\,cm$^{-3}$), the pipeline recovers the injected source at the phase centre with a best-fit DM\,$=$\,150\,pc\,cm$^{-3}$, with individual dispersed pulses reaching a median image-plane S/N of $\approx$52. When integrated into CyberEther, the core processing requires approximately 63\,ms per integration compared with a native integration time of approximately 98\,ms, demonstrating that the core science-processing operations can be completed within the available integration time. The present implementation operates on recorded visibility data and relies on a temporal mean and MAD noise estimate precomputed over the full observation. For live operation, these quantities will instead need to be estimated from data available up to the current integration.

\vspace{2pt}
\noindent\textbf{Keywords:} SETI; technosignatures; transients; interferometric imaging; GPU; RFI mitigation; dispersion measure; CyberEther; radio astronomy

\vspace{8pt}
\end{minipage}
}]



\iacunnumbered{Acronyms/Abbreviations}
\begin{tabular}{@{}l p{0.68\columnwidth}@{}}
ATA & Allen Telescope Array \\
DM & Dispersion Measure (pc\,cm$^{-3}$) \\
IQRM & Inter-Quartile Range Mitigation \\
RFI & Radio-Frequency Interference \\
S/N & Signal-to-Noise ratio \\
SPW & Spectral Window \\
UVH5 & Ultraviolet HDF5 Visibility format \\
\end{tabular}

\section{Introduction}
Technosignature searches have historically emphasized single-dish or beamformed products derived from time--frequency intensity data~\cite{tarter2011ata,gajjar2021gc, gajjar2022spandak, perez2022wow, sheikh2021blc1}. Interferometric imaging provides complementary spatial information, supporting localization of candidates and discrimination between celestial emission and terrestrial interference~\cite{rampadarath2012vlbi, myburgh2026vla}. Image-domain searches can evaluate many lines of sight simultaneously rather than extracting a single known pixel before temporal analysis.

This paper describes the development and testing of a GPU-accelerated interferometric imaging and image-domain DM search pipeline for ATA correlated UVH5 visibilities, together with its implementation within the CyberEther streaming framework. The goal is to develop a framework capable of processing streaming ATA visibility data in real time. The current implementation streams recorded post-correlation visibility data through the pipeline, allowing us to test the imaging and image-domain DM search while working toward live real-time processing at the ATA. 

In this work the pipeline forms ordinary dirty images; CLEAN/PSF deconvolution is available in the imaging path and has been tested, but it is not used for the DM-search results presented here. RFI mitigation is incorporated using the Interquartile Range Mitigation (IQRM) algorithm, which identifies contaminated time-frequency samples. Channels flagged by IQRM are given zero weight during imaging, excluding their contribution to the dirty image without modifying the underlying visibility amplitudes. We then perform and test a blind image-domain trial-DM search using a controlled dispersed pulsed injection. The dispersion search is performed directly on frequency-resolved images, allowing each image pixel to be searched as an independent line of sight. Development and testing were performed on an NVIDIA DGX Spark system (GB10 GPU) using PyTorch/CUDA, with the CyberEther streaming framework running on the same host.

In a SETI survey context, we envision the imaging pipeline operating alongside beamformed search backends. Beamformed broadband pulsed searches such as SPANDAK~\cite{gajjar2022spandak}, as well as real-time deep-learning approaches such as FRBNN~\cite{ma2025frbnn}, can provide a first-pass detection stage, with promising candidates subsequently examined using interferometric imaging. The imaging pipeline developed here is intended to support this follow-up by retaining spatial information and searching many pixels simultaneously. Ultimately, a candidate trigger could also be used to retrieve the corresponding segment of recorded visibility or image data for further analysis. 

Section~\ref{sec:data} summarizes the ATA test data.
Section~\ref{sec:imaging} describes imaging and RFI mitigation.
Section~\ref{sec:dm} covers the image-domain dispersion search and the memory-efficient ring buffer.
Section~\ref{sec:ce} describes the CyberEther architecture.
Section~\ref{sec:results} presents results.
Section~\ref{sec:discussion} discusses current limitations and future work, and
Section~\ref{sec:conclusions} summarizes the main conclusions.

\section{Allen Telescope Array Data and Injected Signal}
\label{sec:data}
The Allen Telescope Array (ATA) is a radio interferometer located at Hat Creek Radio Observatory in Northern California and operated by the SETI Institute. The array consists of 43 6.1-m antennas, with 28 currently available for observations. Most of these antennas are equipped with wideband periodic feeds sensitive from $\approx$1--14\, GHz, allowing observations across a broad range of radio frequencies. In interferometry, signals from pairs of antennas are correlated to produce complex visibilities that encode both spatial and spectral information about the observed field. These correlated visibilities provide the input to the imaging pipeline described in this work. 

\begin{figure}[htbp]
  \centering
  \includegraphics[width=0.95\columnwidth]{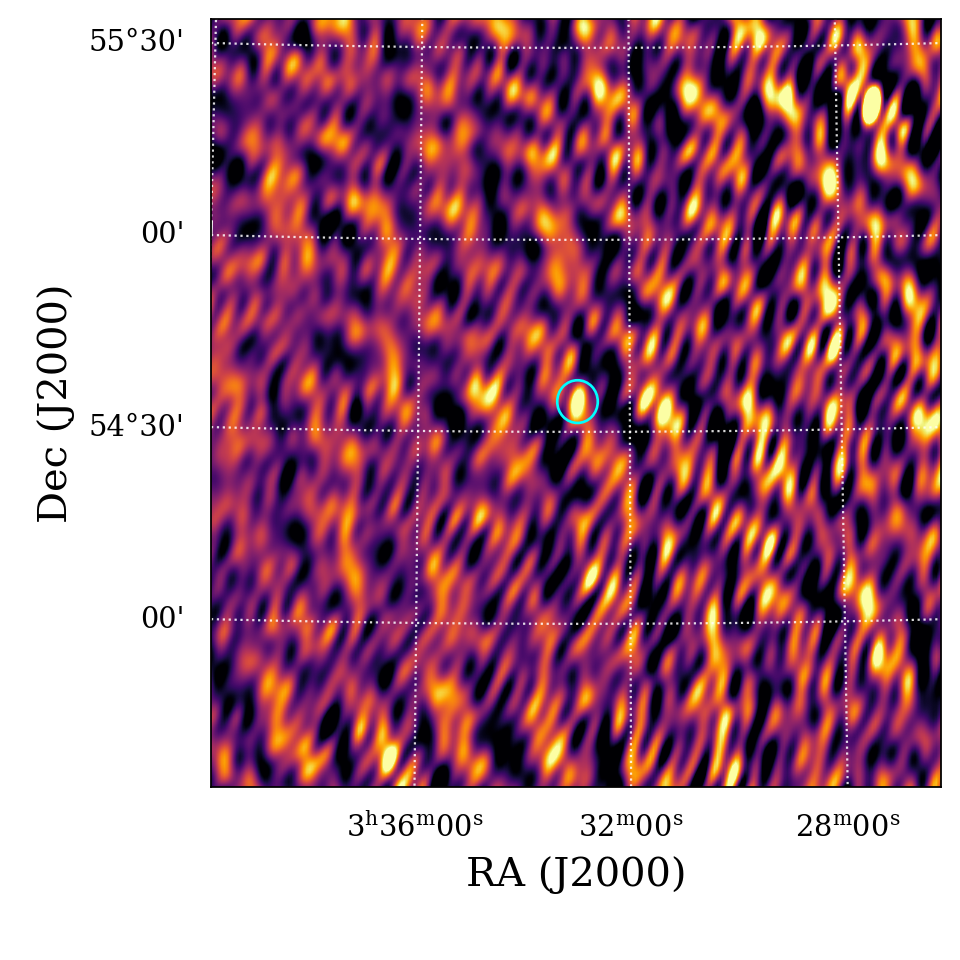}
  \caption{Continuum dirty image of PSR~J0332+5434 from a single integration of SPW~3, produced by the dirty-imaging path described in Section~\ref{sec:imaging} ($128\times 128$ pixel grid, 45\,arcsec cells).}
  \label{fig:j0332}
\end{figure}

The ATA was pointed toward the bright pulsar PSR~J0332+5434 (B0329+54) for a $\approx$10-min observation on 2026 January 21. The observation comprises seven adjacent spectral windows (SPWs) recorded in UVH5 format. Each SPW contains 192 channels with a channel width of 0.5\,MHz, corresponding to 96\,MHz per SPW and a total recorded bandwidth of 672\,MHz. Early continuum dirty-imaging tests used a single SPW (SPW \#3), which had relatively little RFI. The observation contains 6104 integrations with a cadence of
$\Delta t \approx 0.09829$\,s ($\approx 98.29$\,ms), corresponding to a total duration of approximately 10\,min. The visibility data contain 406 baseline products, of which 378 are cross-correlations used for imaging; the 28 auto-correlations are excluded. Dirty images are formed independently for each integration on a $128\times 128$ pixel grid with a cell size of 45\,arcsec, producing a time series of images across the observation. In this imaging time series, PSR~J0332+5434 appears periodically, consistent with its $P\approx 0.7145$\,s pulse period. Fig.~\ref{fig:j0332}) shows an example integration in which the pulsar is detected. 

The DM of PSR~J0332+5434 is low, however, so the in-band dispersive sweep across a single 96\,MHz SPW is only a small fraction of an integration and does not provide a strong test of image-domain DM search. For the multi-band results below, we therefore use the first three SPWs and inject a bright, strongly dispersed pulsed signal. Together, these three SPWs provide 576 contiguous frequency channels spanning  287.5\,MHz, from 1.164 to 1.452\,GHz, on the same time and baseline grid as above; the remaining four SPWs are reserved for future wider-bandwidth tests. 

A periodic dispersed pulsed source with DM\,$=$\,150\,pc\,cm$^{-3}$ is injected into the correlated visibilities at the phase center. The injection is performed directly in the visibility domain: at each selected (time,\,frequency,\,polarization) sample, a real-valued complex constant of amplitude $10^{6}$ in the native, uncalibrated visibility units is added to every baseline product, with zero imaginary part. No target input S/N is specified. Because the added phasor is identical across baselines, the corresponding dirty-image response peaks at the phase centre (pixel $(64,64)$ on the $128\times 128$ grid used here). Each pulse has an intrinsic width of one integration, $\Delta t \approx 98.29$\,ms. Pulses are injected every 100 integrations, corresponding to a period of $100\Delta t\approx 9.829$\,s. After image-domain dedispersion, the injected pulses have a median spatial image-plane S/N of $\approx$52, measured relative to the spatial RMS of source-free pixels in the corresponding DM\,$=$\,150\,pc\,cm$^{-3}$ dedispersed images. 

Frequency-dependent arrival times follow the cold-plasma dispersion law. A delay is computed independently for each of the 576 native frequency channels (not per subband). Every channel frequency $\nu$ is compared to the same global reference, the highest channel centre across all three SPWs,
$\nu_{\mathrm{high}}=1.4515$\,GHz:
\begin{equation}
\tau(\nu)=4.148808\times 10^{-3}\,\mathrm{DM}\,
\left(\nu_{\mathrm{GHz}}^{-2}-\nu_{\mathrm{high}}^{-2}\right)\ \mathrm{s}.
\end{equation}
Thus $\tau(\nu_{\mathrm{high}})=0$, and lower-frequency channels are delayed relative to $\nu_{\mathrm{high}}$. Delays are evaluated on the concatenated 576-channel frequency axis once and then applied within each SPW file, so the three bands therefore share a common dispersion reference rather than independent per-file references. Continuous delays are rounded to the nearest integer number of integrations before injection. For DM\,$=$\,150\,pc\,cm$^{-3}$, the
largest delay is that of the lowest channel centre, $\nu_{\mathrm{low}}=1.164$\,GHz: $\tau(\nu_{\mathrm{low}})\approx 0.164$\,s ($\approx 2$ integrations). 


\section{Interferometric Imaging Pipeline}
\label{sec:imaging}

\subsection{CPU reference imaging}
Stokes-I visibilities are formed from the dual-polarization products, and autocorrelations are excluded. The complex visibilities are then gridded onto the $uv$ plane using nearest-neighbor gridding, in which each visibility is assigned to the nearest $uv$-grid cell. Hermitian conjugate visibilities are added at the corresponding negative $(u,v)$ coordinates. Contributions falling within the same grid cell are accumulated and normalized by the summed imaging weights in that cell. The resulting $uv$ grid is inverse-Fourier transformed to form a dirty image. The same procedure is used to form either a continuum image or frequency-resolved subband images. Convolutional gridding is being investigated as a future improvement to the imaging path. 

A NumPy/CPU implementation provides the reference images used to check the GPU imager. The images produced by the GPU implementation are compared directly with the corresponding CPU images to verify that the two implementations reproduce the same imaging calculation with floating-point precision. 

\subsection{GPU imaging}
The production GPU imager is a PyTorch/CUDA implementation of the same dirty imaging procedure as the CPU reference, and is the imager used in CyberEther. On the GPU, weighted visibilities and their Hermitian conjugates are accumulated onto a complex $uv$ grid; an inverse two-dimensional FFT (\texttt{torch.fft.ifft2}) yields the dirty image. On a single 96\,MHz SPW, end-to-end continuum imaging (one dirty image per integration, including display mapping) takes $\approx$2.9\,ms per integration versus the native $\approx$98\,ms cadence.

\subsection{RFI mitigation with IQRM}
\label{sec:iqrm}
RFI flagging follows IQRM~\cite{morello2022iqrm}. IQRM is used here to identify frequency channels whose visibility statistics are anomalous relative to neighboring channels. The resulting time--frequency mask records, for each integration and frequency channel, whether that channel has been identified as contaminated by RFI. During imaging, flagged channels are excluded by setting their imaging weights to zero, while unflagged channels continue to contribute normally. Mask generation is performed once, offline, on the concatenated three-SPW visibility set:
(1)~form Stokes~I;
(2)~build a time--frequency dynamic spectrum by incoherently summing absolute visibility amplitudes across cross-baselines (\texttt{sum\_abs});
(3)~concatenate the three SPWs in frequency;
(4)~within blocks of 1000 integrations, reduce each block to a per-channel spectral statistic---here, the temporal standard deviation---and apply IQRM neighbor comparisons with threshold~3.0 and radius of order 10\% of the channel count (radius\,58 for 576 channels);
(5) produce a boolean mask $\mathrm{tf_{mask}}[t,f]$.

For each integration $t$, the corresponding row $\mathrm{tf_{mask}}[t,:]$ is applied during imaging: channels marked as RFI are assigned zero weight across all baselines for that integration.
The visibility amplitudes are not rewritten, and the UVH5 files are not modified. On the three-SPW test set, the joint $\mathrm{tf_{mask}}$ occupancy is approximately 29.1\% of time--frequency samples, and 246 of 576 channels are flagged at least once.

\section{Image-Domain Dispersion Search}
\label{sec:dm}

\subsection{Frequency-resolved imaging}
The 576 channels are partitioned into 48 subbands of 12 channels ($\approx 6$\,MHz) each. Every integration produces a stack $I_{\mathrm{sb}}[48,128,128]$ of independently dirty-imaged subband planes, using the same GPU imager and IQRM effective weights as in Section~\ref{sec:imaging}.

\subsection{Temporal background subtraction}
Before searching for dispersed pulses, we remove the time-averaged image from each frequency subband. This suppresses sources and other image structure that are present throughout the observation, making short-duration changes in the images easier to identify. For each subband, the mean intensity of every image pixel is calculated over the full 10-min observation and subtracted from that pixel at each integration: 
\begin{equation}
\begin{aligned}
\mu_\nu(x,y)
&=\langle I_\nu(t,x,y)\rangle_t,\\
I'_\nu(t,x,y)
&=I_\nu(t,x,y)-\mu_\nu(x,y).
\end{aligned}
\end{equation}
Here, $I_\nu(t,x,y)$ is the dirty image at subband $\nu$ and integration $t$ at pixel $(x,y)$, $\mu_\nu(x,y)$ is its time-averaged image, and $I'_\nu(t,x,y)$ is the mean-subtracted image used in the DM search. A source with approximately constant brightness is therefore largely removed, whereas a short-duration pulse remains as a deviation from the temporal mean. 

This subtraction is performed to isolate time-variable emission before dedispersion: it is not part of the dispersion correction itself. For the tests presented here, the mean image is calculated from all 6104 integrations before running the DM search. This is suitable for the recorded dataset used here but cannot be done in the same way on a live stream because future integrations are not yet available. A real-time implementation will therefore require a running estimate of the backgroynd using only previously received data. 

\subsection{IQRM-informed subband masking}
\label{sec:mask}
After temporal background subtraction, the precomputed IQRM flag row for each integration $t$, $\mathrm{tf_{mask}}[t,:]$, is grouped into 48 frequency subbands of 12 channels. If any channel in a subband is flagged, that entire $128\times 128$ subband plane is excluded from the trial-DM combination for that integration. The decision is made independently for each subband and integration time and applies to all image pixels; a subband may be included at some integrations and excluded at others. In this data set, 20 of the 48 subbands are excluded at every integration. 

IQRM flags are used differently during imaging and during DM search. During imaging, individual flagged channels are assigned zero weight, while the remaining unflagged channels within the same subband can still contribute to the dirty image. For the DM search, we adopt a more conservative criterion: if any of the 12 channels in a subband is flagged at a given integration, the entire subband image is excluded from the DM combination at that integration. Both operations use the same precomputed  IQRM $\mathrm{tf_{mask}}$. 

We compared this conservative rule with two less aggressive alternatives: (i) keeping a subband whenever at least one of its 12 channels was unflagged, and (ii) weighting each subband by the fraction of unflagged channels $N_{\mathrm{use}}/12$). Both alternatives retain much more data ($\approx 89.7\%$ of subband--time samples, versus $\approx 50.3\%$ for the conservative rule), but on this data set, they let RFI dominate the blind search and obscure the injected pulse. We therefore use the conservative any-channel rejection for all results below.

\subsection{Blind DM and spatial search}
\label{sec:blind}
We search trial dispersion measures
$\mathrm{DM}=10,20,\ldots,200$\,pc\,cm$^{-3}$ (20 trials). Subband delays are computed using the same cold-plasma relation described in Section~\ref{sec:data}, evaluated at each subband center frequency rather than at the native channel centers used for injection. Continuous delays are rounded to the nearest integer number of integrations, $\mathrm{round}(\tau/\Delta t)$. After applying these integer time shifts, the subband images are averaged over the usable subbands at each pixel, ignoring NaN-valued planes, to form the dedispersion cube $S(\mathrm{DM},t,x,y)$.

This is image-domain dedispersion: every pixel is searched at every trial DM, rather than selecting a known sky position before dedispersion. Candidates are ranked by a signal-to-noise ratio computed from the full dedispersed time series at each pixel and DM, where MAD is the median absolute deviation,
\begin{equation}
\mathrm{S/N}=\frac{\max(S)-\mathrm{median}(S)}{1.4826\,\mathrm{MAD}(S)}.
\end{equation}
For the tests presented here, the median and MAD are calculated from the full dedispersed time series after all integrations have been processed. This form of the detection statistics therefore requires the complete observation and cannot be evaluated in the same way as each integration arrives in a live data stream. A real-time implementation will instead require a noise estimate based on previously received data. On the three-SPW test set, the blind search recovers the injection at the phase center (pixel $(64,64)$) with best-fit DM\,$=$\,150\,pc\,cm$^{-3}$ and a temporal MAD $\mathrm{S/N}\approx 286$; neighboring trials at 140 and 160\,pc\,cm$^{-3}$ give temporal MAD $\mathrm{S/N}\approx 272$. This differs from the median image-plane S/N of $\approx$52 reported in Section~\ref{sec:data}: the image-plane S/N compares the source with the spatial RMS of source-free pixels in an individual dedispersed image, whereas the temporal MAD S/N compares the strongest pulse with the variation over time at the same pixel. The ranking selects the strongest individual pulse in the injected periodic train.

\subsection{Ring-buffer implementation}
\label{sec:ring}
Storing the full subband cube ($48\times 6104\times 128\times 128$ float32) requires approximately 17.4\,GB. For the present DM grid, however, the maximum integer delay is only $d_{\max}=2$ integrations, so only a short history of recent subband images is required. We use a ring buffer of depth $R=8$ ($R>d_{\max}+1$), which is sufficient to form all trial-DM alignments without retaining the full frequency$\times$time$\times$image cube in memory. This reduces the resident image history to $\approx 30$\,MB. The
trial-DM image combination was $\approx$ $5.2\times$ faster when operating on the ring buffer than when scanning the full stored cube.

The buffering strategy is inspired by the streaming image-set and ring-buffer concepts in STRIDE~\cite{dipietrantonio2026stride}. Here, the ring serves as a short-history buffer for forming dispersion-aligned image combinations and is not a full implementation of STRIDE. 

\section{CyberEther Streaming Architecture}
\label{sec:ce}

CyberEther is an open-source framework for building real-time signal-processing pipelines as a visual flowgraph of connected modules, with heterogeneous execution across CPU and GPU backends~\cite{cyberether}. Each module separates core processing logic, memory buffers for inputs and outputs, and device-specific drivers, so the same pipeline stage can target different hardware backends. In this work, we use it as the streaming host for post-correlation UVH5 replay: the offline science path is expressed as a graph of processing blocks that exchange tensors between stages. GPU-based real-time backends such as BLADE are also being developed for the ATA~\cite{cruz2023blade}, as part of broader efforts toward real-time GPU-based signal processing. 

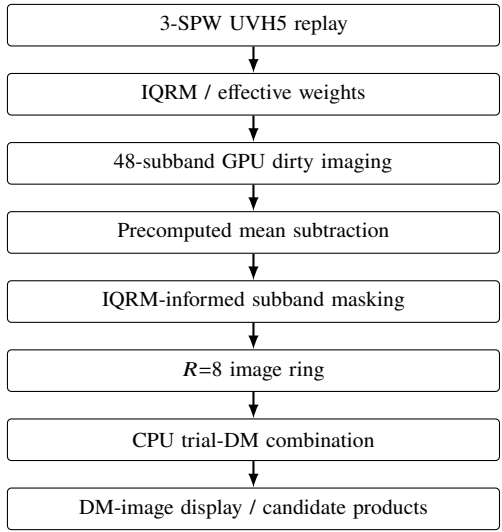
\begin{figure}[htbp]
  \centering
  \begin{tikzpicture}[
    font=\footnotesize,
    node distance=0.35cm,
    stage/.style={
      draw,
      rounded corners=1.5pt,
      align=center,
      text width=0.82\columnwidth,
      inner sep=3pt,
      minimum height=0.55cm
    },
    arr/.style={-latex, thick}
  ]
    \node[stage] (n1) {3-SPW UVH5 replay};
    \node[stage, below=of n1] (n2) {IQRM / effective weights};
    \node[stage, below=of n2] (n3) {48-subband GPU dirty imaging};
    \node[stage, below=of n3] (n4) {Precomputed mean subtraction};
    \node[stage, below=of n4] (n5) {IQRM-informed subband masking};
    \node[stage, below=of n5] (n6) {$R{=}8$ image ring};
    \node[stage, below=of n6] (n7) {CPU trial-DM combination};
    \node[stage, below=of n7] (n8) {DM-image display / candidate products};

    \foreach \a/\b in {n1/n2,n2/n3,n3/n4,n4/n5,n5/n6,n6/n7,n7/n8}
      \draw[arr] (\a) -- (\b);
  \end{tikzpicture}
  \caption{CyberEther replay graph for the three-SPW pipeline.}
  \label{fig:cegraph}
\end{figure}

The CyberEther replay graph used in this work is shown in Fig.~\ref{fig:cegraph}. Only dirty imaging runs on the GPU. Replay, IQRM weight application, mean subtraction, IQRM-informed subband masking, ring management, and trial-DM combination remain CPU stages in the present pipeline. Visibilities bypass the IQRM node; only weights are modified.

\section{Results}
\label{sec:results}
\begin{figure}[t]
  \centering
  \includegraphics[
    width=\columnwidth
  ]{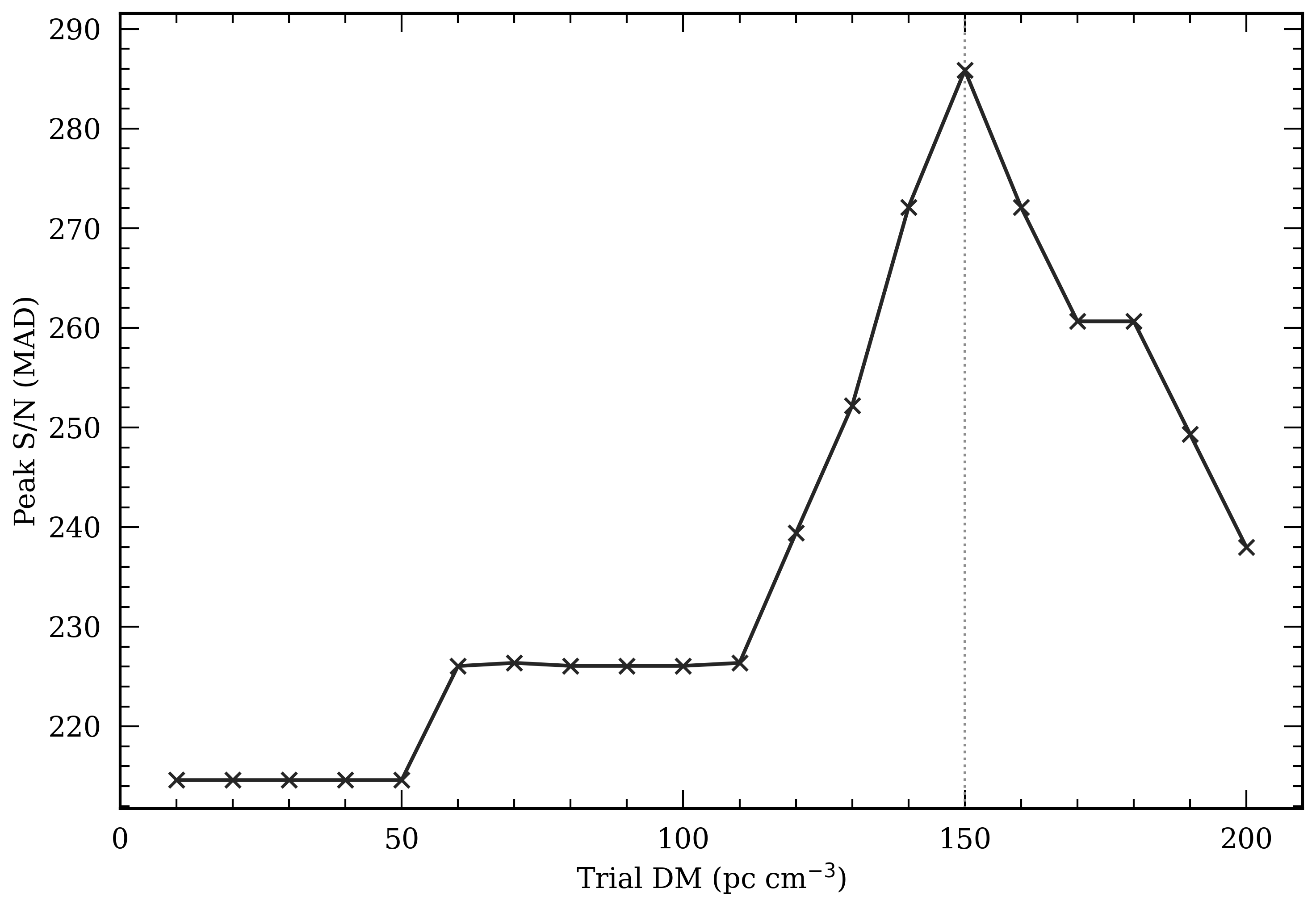}
  \caption{Phase-centre S/N versus trial DM for the injected signal.}
  \label{fig:snrdm}
\end{figure}

\begin{figure}[t]
  \includegraphics[
    width=\columnwidth]{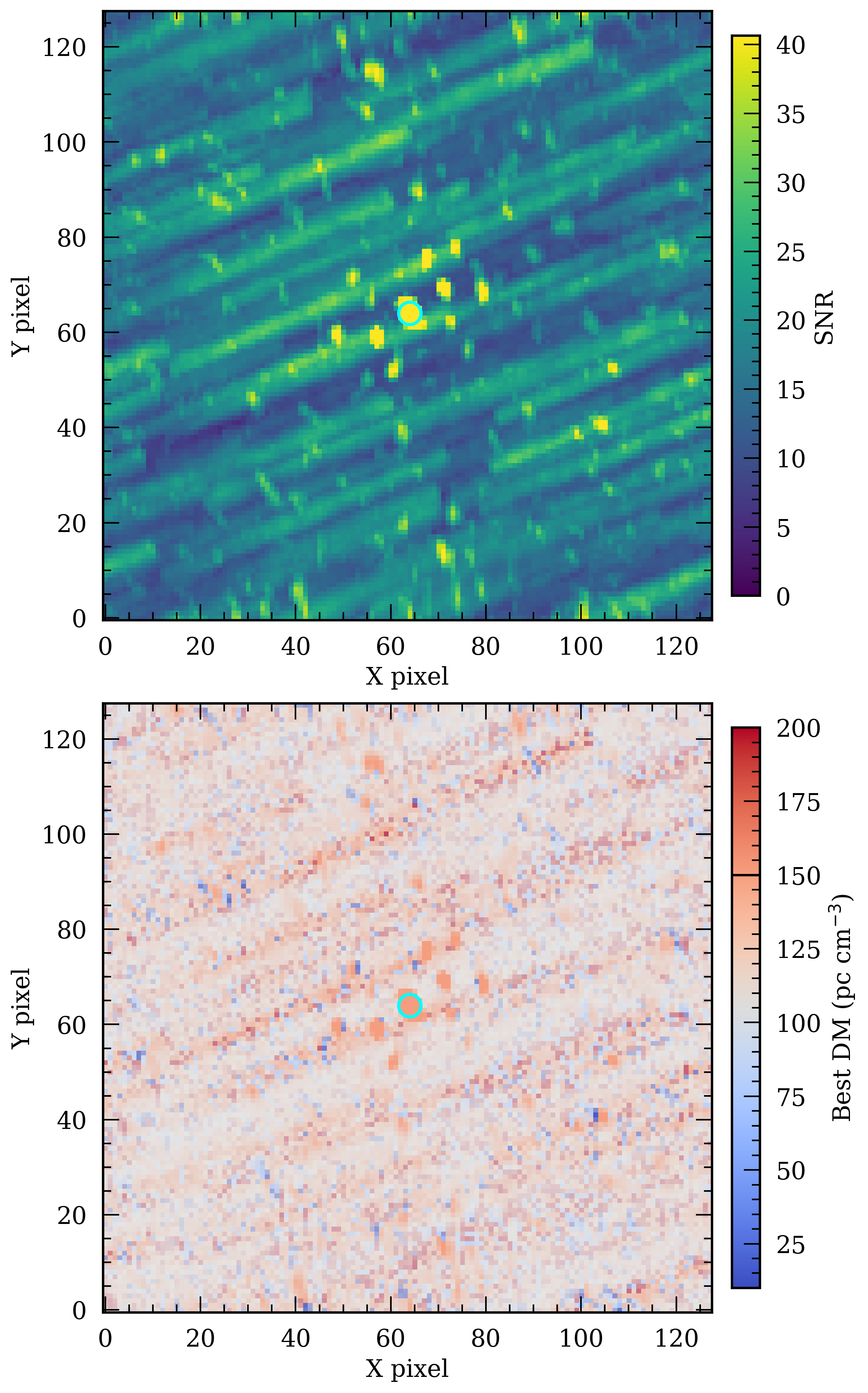}
  \caption{Blind search maps: peak S/N (top) and best-fit DM (bottom), per pixel.}
  \label{fig:scan}
\end{figure}

As described in Section~\ref{sec:blind}, the search recovers the injected source at the phase-centre pixel $(64,64)$ with
$\mathrm{DM}=150$\,pc\,cm$^{-3}$ and $\mathrm{S/N}\approx 286$. Neighboring trials at 140 and 160\,pc\,cm$^{-3}$ give $\mathrm{S/N}\approx 272$. Fig.~\ref{fig:snrdm} shows the phase-centre S/N--DM curve; Fig.~\ref{fig:scan} shows the blind pixel search maps. The search also produces a candidate list containing the pixel location, best-fit DM, pulse time, and S/N of the highest-ranked events. The same injected signal is recovered when the pipeline is executed within CyberEther as described in Section~\ref{sec:ce}. CyberEther displays the dedispersed image each integration, while candidate ranking across image pixels, trial DMs, and time is performed separately offline.

\begin{table}[htbp]
\caption{Approximate per-integration processing times for the CyberEther science path.}
\label{tab:timing}
\centering
\fontsize{8}{9}\selectfont
\begin{tabular}{@{}lr@{}}
\toprule
Stage & Time (ms/integration) \\
\midrule
Replay / I/O & 9.66 \\
IQRM apply & 0.09 \\
GPU subband imaging & 33.05 \\
Mean subtraction & 0.18 \\
Subband masking & 0.18 \\
Ring buffer & 0.10 \\
CPU DM combination & 20.08 \\
\midrule
Science total & $\approx$63.3 \\
Native cadence & 98.29 \\
\bottomrule
\end{tabular}
\end{table}

The core science stages require $\approx 63.3$\,ms per integration, which is below the native $\Delta t\approx 98.29$\,ms cadence (Table~\ref{tab:timing}). Writing the full dedispersed $S$ product adds $\approx 50$--$55$\,ms/integration, increasing the total processing time to $\approx 124$\,ms/integration. We therefore distinguish the science processing time from output I/O and do not treat the I/O-inclusive run as real-time.

\section{Discussion and Current Limitations}
\label{sec:discussion}
The present search uses a temporal mean and MAD noise estimate computed from the full observation. For live operation, these quantities will need to be estimated using only data up to the current integration. The system also remains a post-correlation pipeline operating on recorded UVH5 data rather than a live ATA correlator backend. Candidate ranking is currently performed offline after replay rather than during streaming. Periodicity is not included in the present candidate ranking stage, but we have developed tools that can browse dedispersed time series and run FFT/fold follow-ups. 

Additionally, GPU subband imaging is divided into 48 relatively small operations, with repeated data transfers, so GPU utilization is not yet optimal. Evaluation so far has emphasized a controlled injected signal rather than blind searches on real data. The conservative any-channel subband mask worked well on this dataset, but milder or spatially dependent RFI treatments remain to be explored.

Future work includes batching GPU subband imaging to improve occupancy
beyond the present 48 serial subband operations, running background and noise estimators for live processing, performing candidate detection during streaming without large $S$ data products, and periodicity scoring on dedispersed per-pixel time series. We will also test the full image-domain DM search on individual pulses from known, more highly dispersed pulsars, including B1933+16 (DM\,$\sim$\,158\,pc\,cm$^{-3}$). Injection tests spanning a range of DMs and S/N ratios will additionally be used to characterize the recovery performance of the image-domain search. 

\section{Conclusions and Future Work}
\label{sec:conclusions}
We have developed a GPU-accelerated dirty imaging and image-domain DM search pipeline for ATA UVH5 visibilities, with IQRM-informed weighting, conservative IQRM-informed subband masking, a
short-history ring buffer, and a CyberEther replay graph. On a three-SPW injected dataset, the pipeline recovers the dispersed pulsed source
at the phase centre with DM\,$=$\,150\,pc\,cm$^{-3}$ and S/N\,$\approx 286$. The core processing finishes within one integration ($\approx 63$\,ms versus the $\approx 98$\,ms cadence), but writing the large
dedispersed $S$ product does not.

Near-term work includes batched GPU subband imaging, migrating more of the pipeline from CPU to GPU, background and noise estimators for live processing, candidate detection during streaming without large $S$ products, periodicity scoring of dedispersed per-pixel time series, and tests on real data.

For SETI survey use, we envision pairing beamformed first-pass searches with the imaging pipeline developed here. Candidate events identified in
beamformed products could trigger extraction of the corresponding visibility or image data segment for image-domain localization and confirmation.

Longer-term, we aim to connect this post-correlation imaging and search pipeline with complementary real-time voltage-domain streaming work currently under development, so that visibility formation and image-domain processing can operate together in a live ATA system~\cite{cruz2023blade}. The goal is a continuous real-time science capability for survey observing.

\iacunnumbered{Acknowledgements}
We thank the staff scientists and engineers at the ATA for keeping our antennas working around the clock, supporting science data products, CyberEther development, and GPU computing resources used for this work.



\end{document}